\documentclass[epj]{svjour}
\usepackage{graphics}
\begin{document}
\title{First Measurement of the $\Sigma$ Beam Asymmetry in $\eta^{\prime}$ Photoproduction off the Proton near Threshold}
\author{P. Levi Sandri\inst{1} \thanks{email: paolo.levisandri@lnf.infn.it} \and G. Mandaglio\inst{2,3} \thanks{email: gmandaglio@unime.it}\and V. De Leo\inst{4} \and O. Bartalini\inst{5,6} \and V. Bellini\inst{7,8} \and J.-P. Bocquet\inst{9} \and M. Capogni\inst{5,6}\thanks{\emph{Present address:} ENEA - C.R. Casaccia, Istituto Nazionale di Metrologia delle Radiazioni Ionizzanti, Via Anguillarese, 301 I-00123 Roma, Italy} \and F. Curciarello\inst{2,8}  \and J.-P. Didelez\inst{10} \and A. D'Angelo\inst{5,6} \and R. Di Salvo\inst{5} \and A. Fantini\inst{5,6} \and D. Franco\inst{5,6}\thanks{\emph{Present address:} IPNL - 43, Bd du 11 Novembre 1918, Fr69622 Villeurbanne Cedex, France} \and G. Gervino\inst{11,12} \and F. Ghio\inst{13,14} \and  \and B. Girolami\inst{13,14} \and A. Giusa\inst{7,8} \and A. Lapik\inst{15} \and A. Lleres\inst{9} \and F. Mammoliti\inst{7,8} \and M. Manganaro\inst{2,8}\thanks{\emph{Present address:} Universidad de La Laguna, Instituto de Astrof\'isica de Canarias, E-38205 La Laguna, Tenerife, Spain} \and D. Moricciani\inst{5} \and A. Mushkarenkov\inst{15} \and V. Nedorezov\inst{15} \and C. Randieri\inst{7,8} \and D. Rebreyend\inst{9} \and N. Rudnev\inst{15} \and G. Russo\inst{7,8} \and C. Schaerf\inst{5,6} \and M.-L. Sperduto\inst{7,8} \and M.-C. Sutera\inst{8} \and A. Turinge\inst{15} \and V. Vegna\inst{5,6}\thanks{\emph{Present address:} Universit\"{a}t Bonn, Physikalisches Institut - Nu\ss allee 12,  Bonn, D-53115, Germany} \and I. Zonta\inst{5,6}
}                     % Do not remove
\institute{INFN - Laboratori Nazionali di Frascati,  00044 Frascati, Italy 
\and 
Dipartimento di Fisica e di Scienze della Terra, Universit\`a di Messina, 98166 Messina, Italy 
\and 
INFN - Gruppo collegato di Messina, 98166 Messina, Italy  
\and
Dipartimento di Matematica e Fisica dell’Universit\'a ``Roma Tre'', and
INFN Sezione di Roma Tre, Roma, Italy
\and 
INFN - Sezione di Roma Tor Vergata, 00133 Roma, Italy 
\and
Dipartimento di Fisica - Universit\`a degli Studi di Roma Tor Vergata, 00133 Roma, Italy 
\and
Dipartimento di Fisica - Universit\`a degli Studi di Catania, 95123 Catania, Italy
\and
INFN - Sezione di Catania, 95123 Catania, Italy
\and
LPSC, Universit\'{e} Grenoble-Alpes, CNRS/IN2P3, Grenoble IPN, F-38026 Grenoble, France
\and
IN2P3, Institut de Physique Nucl\'eaire, 91406 Orsay, France
\and
INFN- Sezione di Torino, 10125 Torino, Italy
\and
Dipartimento di Fisica Sperimentale - Universit\`a degli Studi di Torino, I-10125 Torino, Italy
\and
INFN - Sezione di Roma, 00185 Roma, Italy
\and
Istituto Superiore di Sanit\`a,  I-00161 Roma, Italy
\and
Institute for Nuclear Research, 60-letiya Oktyabrya prospekt 7a, 117312 Moscow, Russia
}
\date{Received: date / Revised version: date}
\abstract{
The $\Sigma$ beam asymmetry in $\eta^{\prime}$ photoproduction off  the proton was measured at the GrAAL polarised photon beam with incoming photon energies of 1.461 and 1.480 GeV. For both energies the asymmetry as a function of the meson production angle shows a clear structure, more pronounced at the lowest one, with a change of sign around 90$^{\circ}$.  The observed behaviour is compatible with P-wave D-wave (or S-wave F-wave) interference, the closer to threshold the stronger. The results are compared to the existing state-of-the-art calculations that fail to account for the data.
 % end of PACS codes
} %end of abstract
\authorrunning{P. Levi Sandri {\it et al.}}
\titlerunning{First Measurement of the $\Sigma$ Beam Asymmetry in $\eta^{\prime}$ Photoproduction ...}
\maketitle
%

%\section{I Introduction}
The experimental study of nucleon excited states is fundamental for the understanding of its internal structure. Important differences are still observed today between the experimental nucleon spectrum and the predictions of the first Constituent Quark Models (CQM)\cite{Capstick:1986bm,Capstick:1992uc,Capstick:1993kb,Riska:2000gd} but also with the results of recent approaches like lattice QCD calculations\cite{Edwards:2011jj}, Dyson-Schwinger equation of QCD\cite{Roberts:2011cf}, harmonic oscillator CQM\cite{Klempt:2012fy} and hypercentral CQM\cite{Giannini:2015zia}. Recent reports on advances in the experimental studies of the excited nucleon state spectrum can be found in\cite{Crede:2013kia,Burkert:2013dia}

Several states predicted by these models have not been observed ({\it missing resonances}). The nucleon excited states decay strongly with meson emission; therefore meson photoproduction experiments off the nucleon are an ideal way of searching for missing resonances and complement the information obtained with pion-nucleon scattering experiments.

In pseudo-scalar meson photoproduction off the proton ($\gamma + p \rightarrow meson + p$) we have eight possible combinations of spin states. The scattering amplitude is thus described by eight matrix elements, only four of which are independent due to rotational invariance and parity transformations.
With these four complex amplitudes, 16 bilinear products can be constructed, corresponding to 16 observables: the differential cross section, three single polarisation observables and twelve double polarisation observables.
To determine the scattering amplitude thoroughly, the cross section, the
three single polarisation and four appropriately chosen double
polarisation observables must be measured \cite{Keaton:1996pe,Chiang:1996em}. These observables can be expressed in terms of helicity amplitudes and the following relations hold\cite{Drechsel:1998hk,Feuster:1998cj,Saghai:1996hn,Fasano:1992es}:

${d\sigma}/{d\Omega} \sim |H_{1}|^{2} + |H_{2}|^{2} + |H_{3}|^{2} + |H_{4}|^{2}$

$\Sigma \sim Re(H_{1}H_{4}^{*} - H_{2}H_{3}^{*})$

$T \sim Im(H_{1}H_{2}^{*} - H_{3}H_{4}^{*})$

$P \sim Re(H_{1}H_{3}^{*} - H_{2}H_{4}^{*})$ \newline
where ${d\sigma}/{d\Omega}$ is the differential cross section and $\Sigma$, $\it{T}$ and $\it{P}$ are the beam, target and recoil asymmetries respectively. From the above relations one can see that the 
the amplitude phases can not be accessed from the data on differential cross
section alone and that the only source of information to determine them are the polarisation asymmetries.
The combined study of the unpolarised cross section and of the polarisation
asymmetries is critical in order to get access to the
production amplitudes from the experimental data, and the interference among the helicity amplitudes can play a crucial role in revealing subtle effects\cite{Arndt:1989ww}.

The pseudo-scalar nature of the $\eta^\prime$ meson ensures that only $N^*$ resonances contribute to the process. The production threshold at W = 1.896 GeV (corresponding to an incident photon energy of 1.447 GeV for a free proton target) is located just above the so-called $resonance~gap$, where many of the predicted, but so far unobserved, $N^*$ states should be located.

The first data on $\eta^{\prime}$ photoproduction cross section were produced in 1968 
%in a bubble chamber experiment
\cite{ABBHHM:1968aa} 
%using an untagged photon beam
, and confirmed in 1976
% with a streamer chamber setup and tagged photons at DESY
\cite{Struczinski:1975ik}. Over 20 years later, the SAPHIR collaboration\cite{Plotzke:1998ua} reported a more extended measurement, based on 250 events, from which the masses and widths of the dominating $S_{11}$ and $P_{11}$ resonances were extracted.
In more recent years, the CLAS experiment at Jlab and the CB-ELSA-TAPS in Bonn have produced a rich amount of precise total and differential cross section data on the proton\cite{Dugger:2005my,Williams:2009yj,Crede:2009zzb} in the energy region from threshold up to 2.84 GeV.

From the theoretical point of view, four approaches are available in the literature:
a relativistic meson-exchange model of hadronic interactions\cite{Nakayama:2005ts,Huang:2012xj}; %, with t-channel mesonic currents ($\rho$ and $\omega$) and s- and u-channel resonances contributions. This approach was later revisited\cite{Huang:2012xj} performing a combined analysis of $\eta^{\prime}$ production reactions.
 a reggeized model for $\eta$ and $\eta^{\prime}$ photoproduction\cite{Chiang:2002vq}; %where the vector meson exchanges are treated in terms of Regge trajectories to comply with the correct high-energy behaviour.
 a chiral quark-model\cite{Zhong:2011ht} and 
%with the process governed by $S_{11}$(1535) and u-channel background.
 an isobar model\cite{Tryasuchev:2013fua}.
%where a good description of the existing cross section data is obtained by taking into account the contributions of  6 high-angular-momenta heavy resonances alone.

As a consequence of this huge experimental and theoretical effort, it was established that three above-threshold resonances ($S_{11}, P_{11}, P_{13}$), and the four-star sub-threshold $P_{13}$(1720) resonance reproduce best all existing data for the $\eta^{\prime}$ production processes in the resonance-energy region\cite{Huang:2012xj}, and that above 2 GeV, where the process is dominated by the $\rho$ and $\omega$ exchange, the dynamics of $\eta^{\prime}$ photoproduction are similar to those of $\eta$ photoproduction\cite{Crede:2009zzb}.

All the abovementioned state-of-the-art theoretical calculations give a reasonable description of the data. In all cases the authors stress that the cross section data alone are unable to pin down the resonance parameters, while polarisation observables could be very helpful to better determine the partial wave contributions in this reaction and impose more stringent constraints on the parameter values of the different models.

In this letter, we present the first measurement of the single polarisation observable $\Sigma$ for $\eta^{\prime}$ photoproduction off the proton, at the incoming photon energies of 1.461 and 1.480 GeV, obtained with the Compton backscattered photon beam of the GrAAL experiment. 

The GrAAL experiment was located at the European Synchrotron Radiation Facility (ESRF) in Grenoble
(France), where it took data from 1995 to 2008.
A linearly polarised photon beam impinged on a liquid H2 or D2 target, and the final products were detected by the large solid angle detector LAGRAN$\gamma$E (Large Acceptance GRaal-beam Apparatus for Nuclear $\gamma$ Experiments).

The photon beam was produced by the Compton back-scattering of low-energy polarised photons from an
Argon laser, against the 6.03 GeV electrons circulating inside the ESRF storage ring\cite{DAngelo:2000ce}. The UV laser line (3.53 eV) was used to produce a backscattered photon beam, covering the energy range up to 1.5 GeV. A tagging system, located inside the electron ring, provided an event-by-event measurement of the photon beam energy, with a resolution of 16 MeV (FWHM).
Since the electron involved in the Compton scattering is ultra-relativistic, its helicity is conserved in the
process at backward angles, and the outgoing photon retains the polarisation of the incoming laser beam (up to 96\% for the UV laser line). The correlation between photon energy and polarisation is calculated with QED \cite{Babusci:1995ji}. During data taking, the laser beam polarisation was rotated by 90$^{o}$ every 20 minutes approximately, and unpolarised data from the Bremsstrahlung of the electrons off the ESRF residual vacuum were collected as well.

A detailed description of the LAGRAN$\gamma$E apparatus can be found in \cite{Bartalini:2005wx}. For the purpose of this letter
we underline the excellent energy resolution of the BGO electromagnetic calorimeter (${\it Rugby~Ball}$)\cite{LeviSandri:1996tk} where
photons from the  $\eta^{\prime}$ decay chain were measured, and the position and time resolution in the forward direction
(1.5$^{\circ}$ and  2$^{\circ}$ (FWHM) for polar and azimuthal angles, respectively and 300 ps for time-of-flight (TOF)) where the recoil protons
from the photoreaction were detected.

Data were collected during eight different stretches of the GrAAL experiment, from 1998 to 2002. As the threshold for $\eta^{\prime}$ photoproduction off the proton is $E_{th} = 1.447$~GeV, only the periods of measurement performed by using the UV laser line (351 nm wavelength) allow to reach $E_{th}$ and to explore the behaviour of the asymmetry as a function of the photon energy up to 1.5 GeV. The $\eta^{\prime}$ mesons were identified via $\gamma\gamma$, $\pi^{0}\pi^{0}\eta$ and $\pi^{+}\pi^{-}\eta$ decay modes and by requiring the fulfilment of the two-body kinematics for the recoil proton.

\begin{figure}[htbp]
 \begin{center}
\resizebox{0.48\textwidth}{!}{\includegraphics{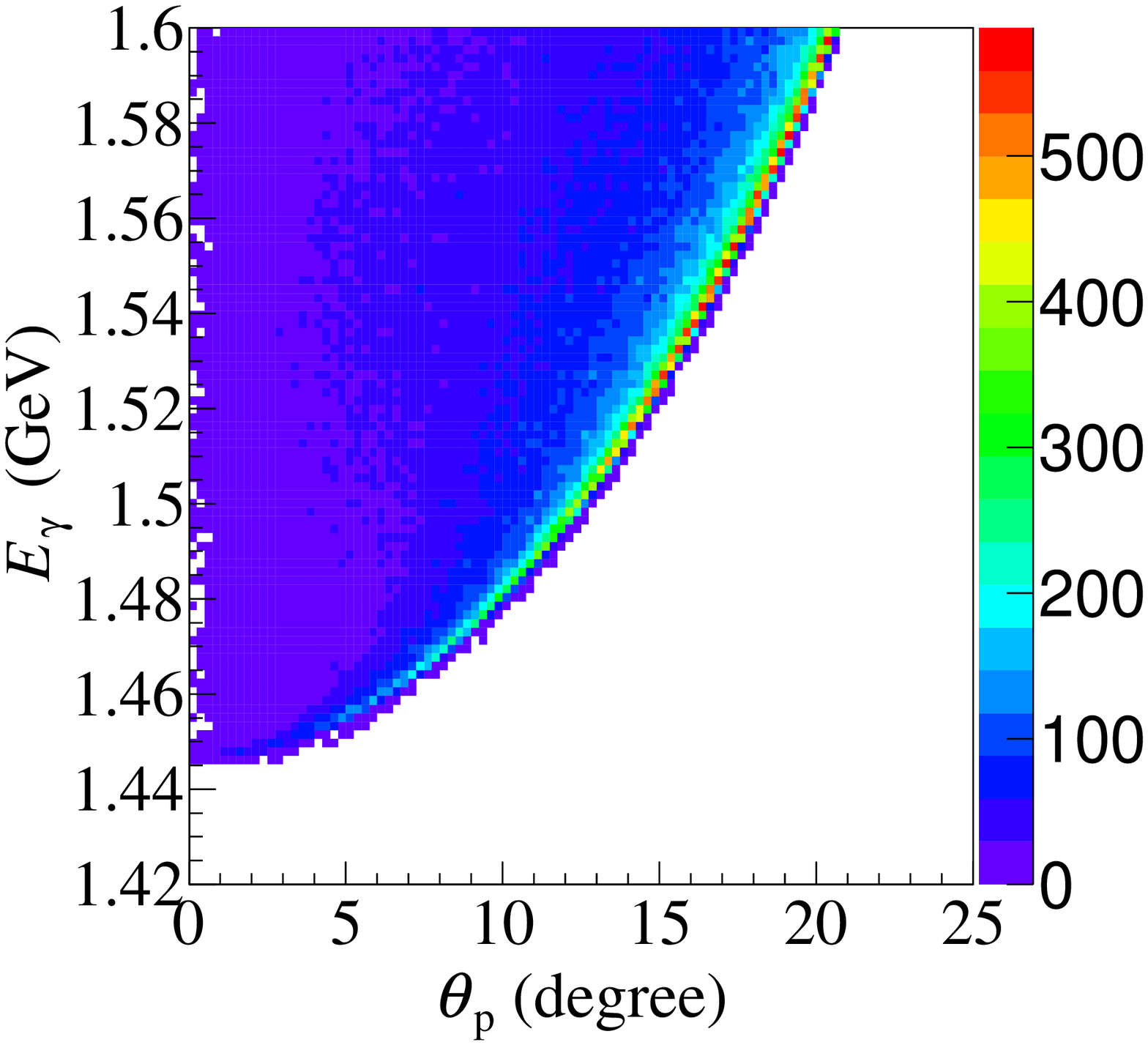}\includegraphics{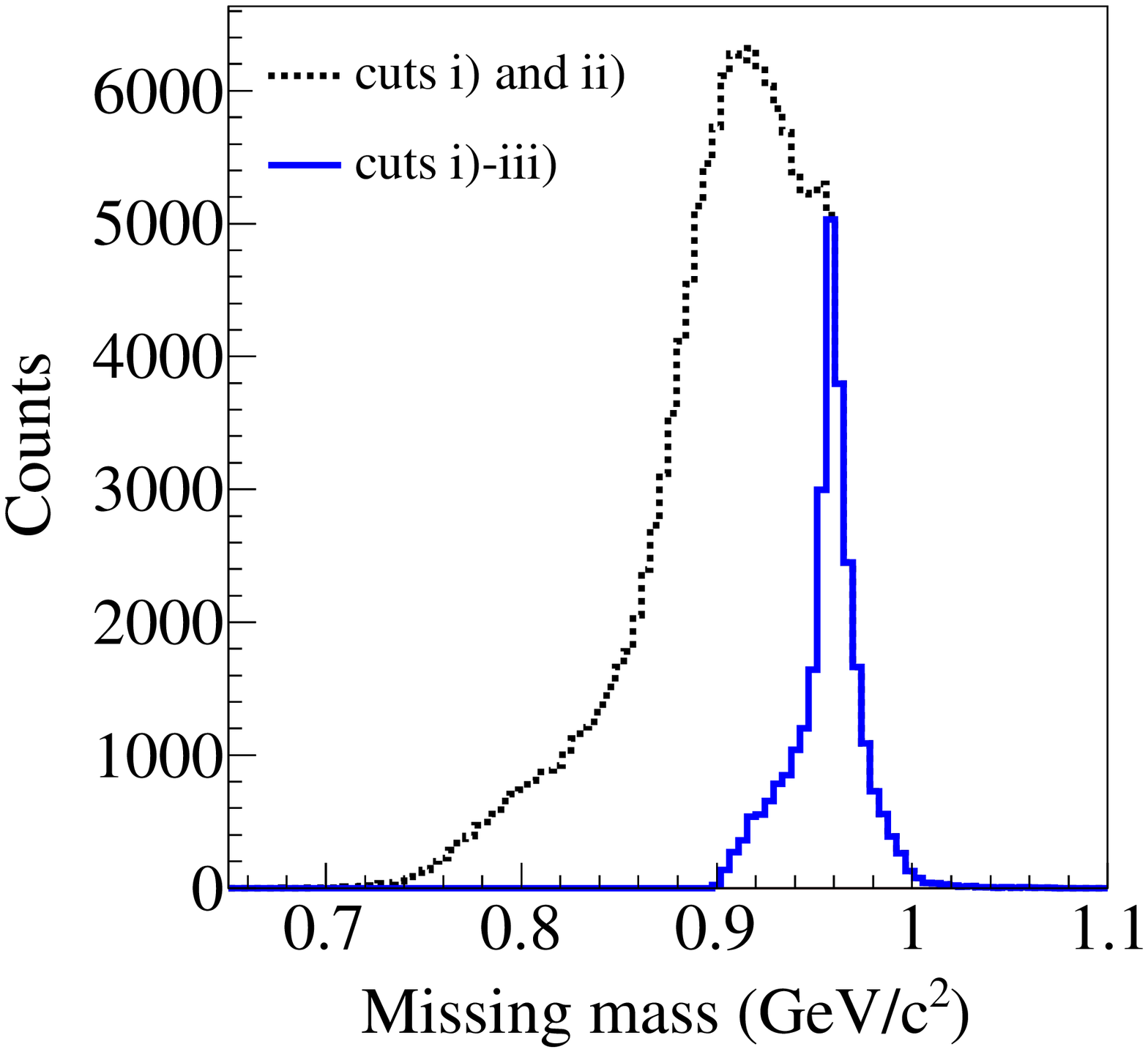}}
  \put(-183,50){(a)} \put(-95,50){(b)}\\
\resizebox{0.50\textwidth}{!}{\includegraphics{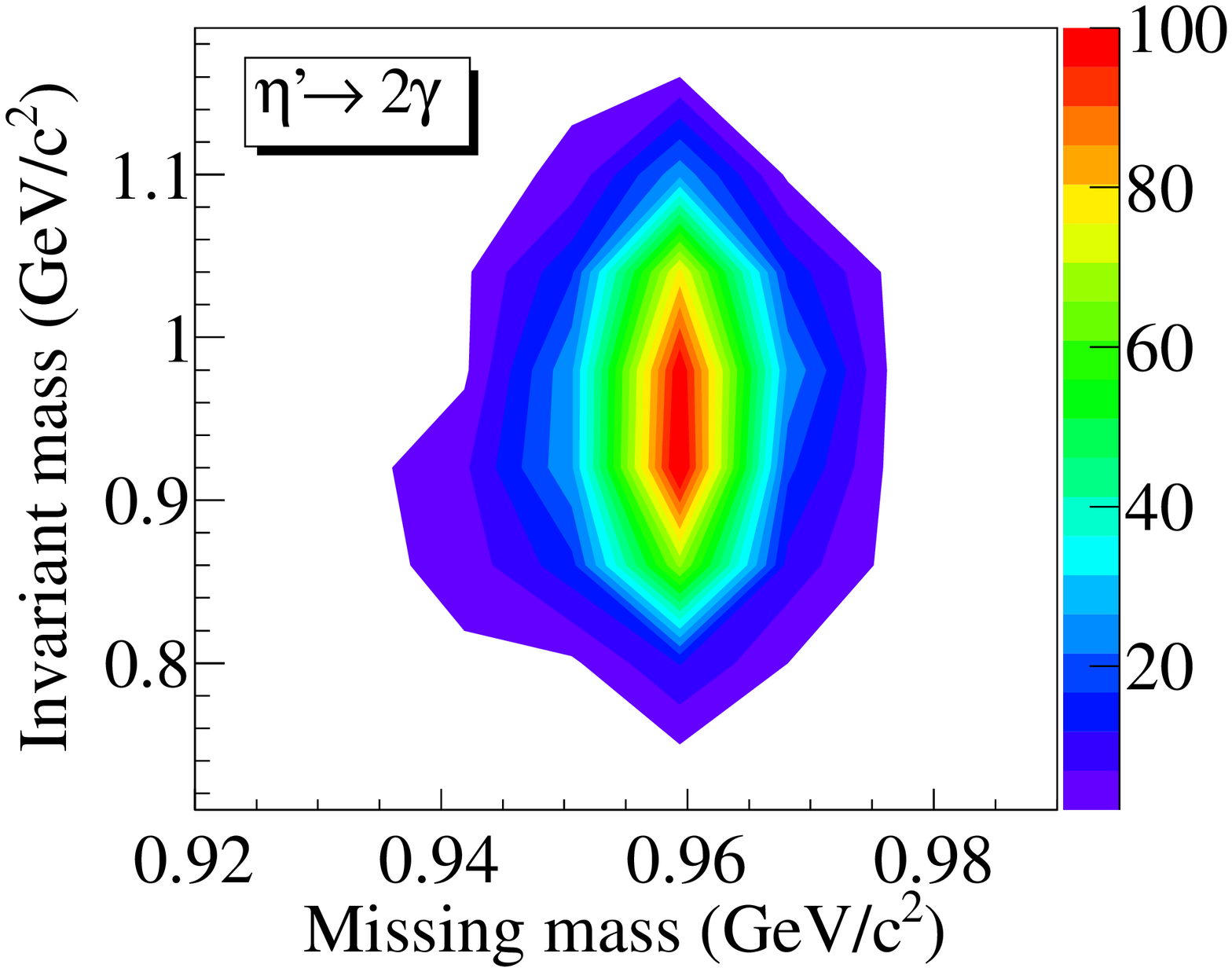}\includegraphics{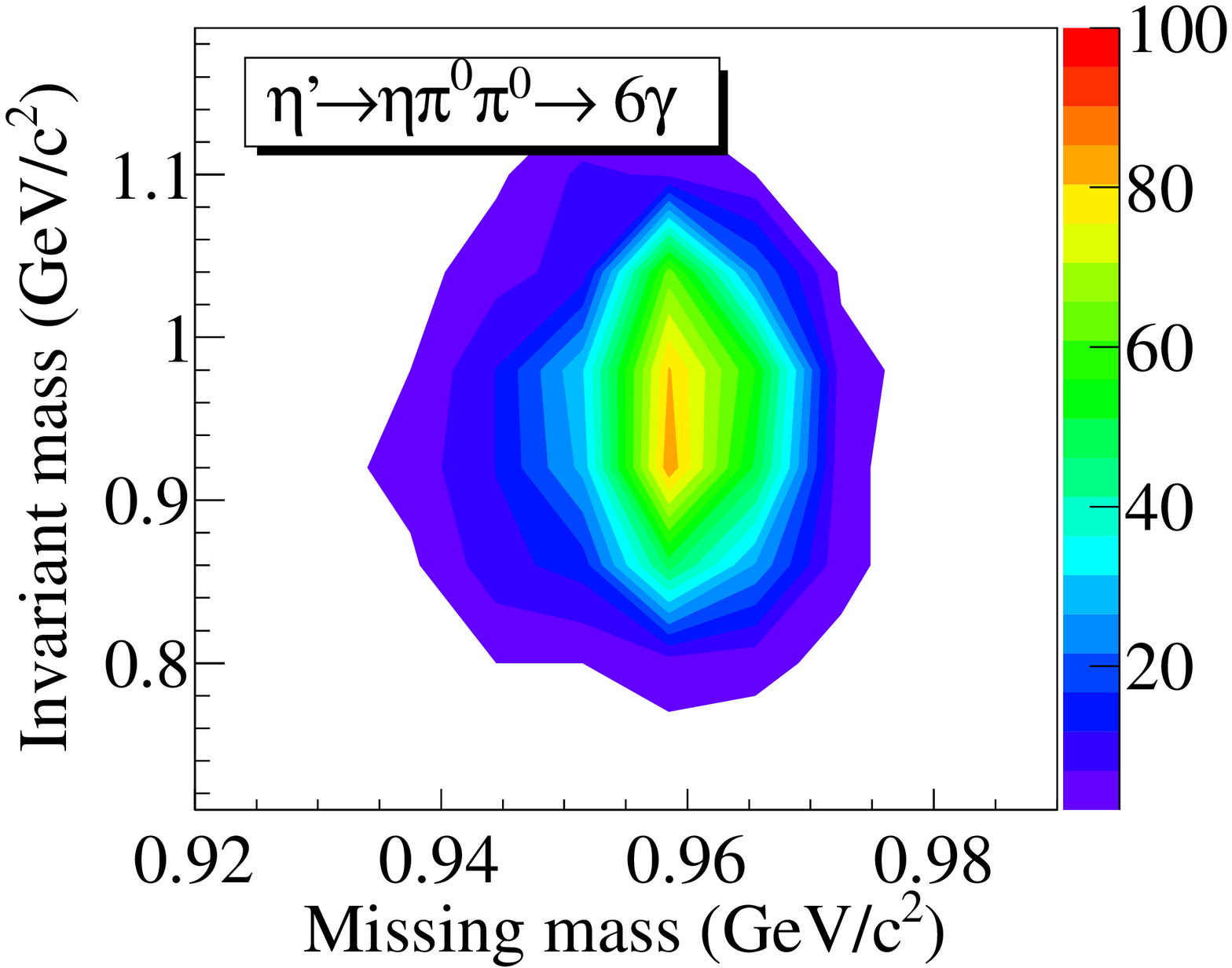}}
\put(-233,50){(c)} \put(-105,50){(d)}\\
\resizebox{0.50\textwidth}{!}{\includegraphics{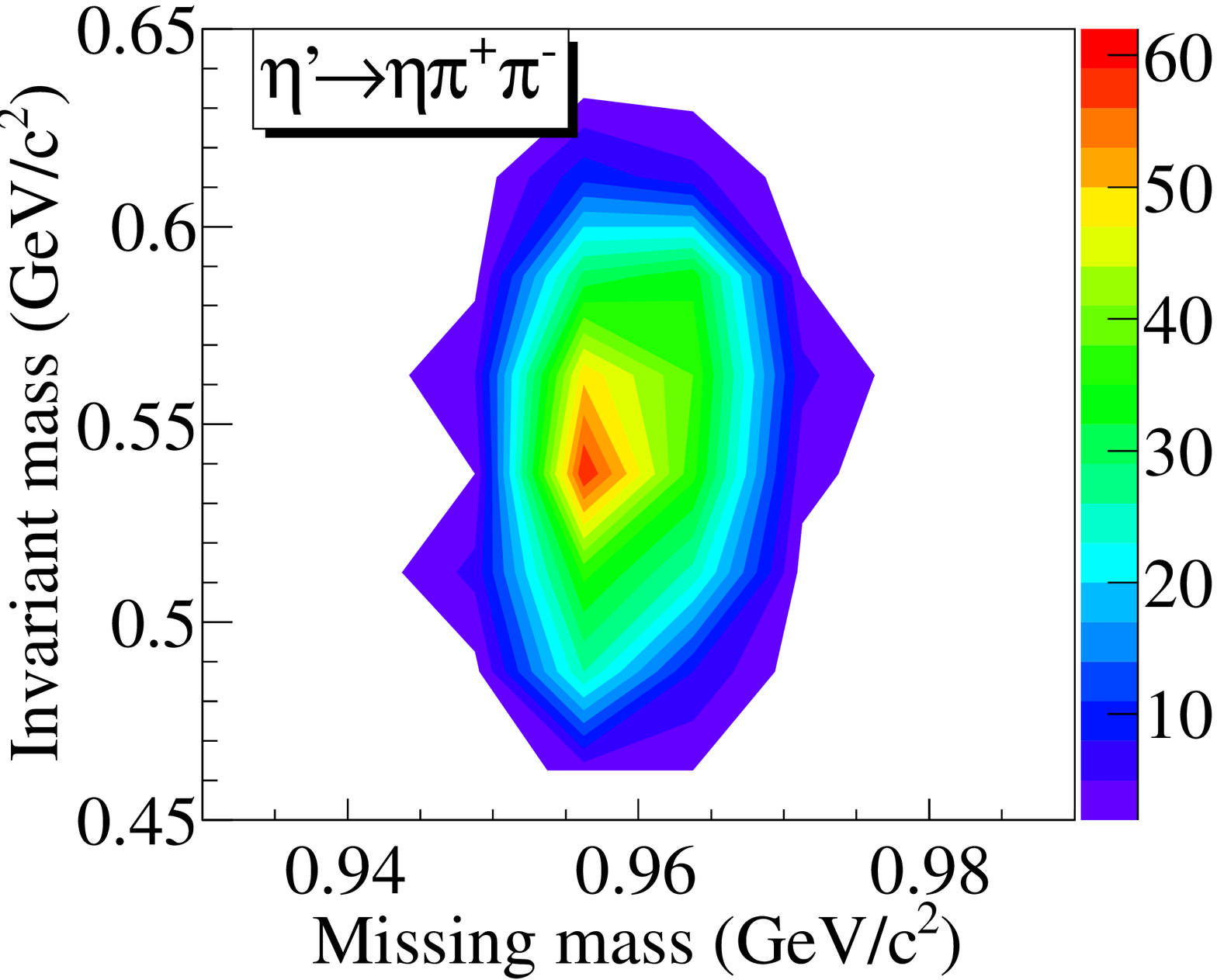}\includegraphics{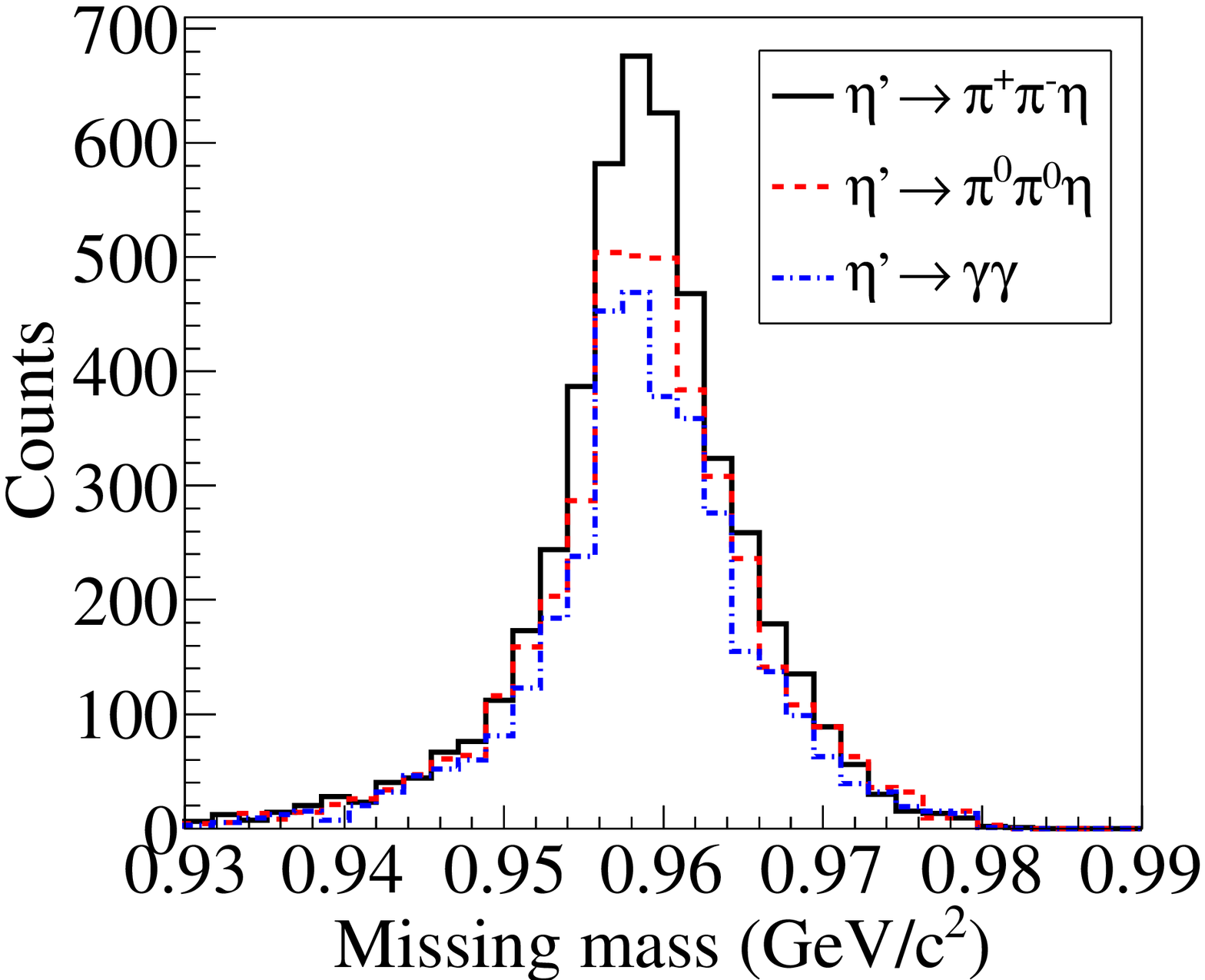}}
\put(-233,50){(e)} \put(-105,50){(f)}
\caption{(Colour online) Panel a: energy of photon beam $\it{vs.}$ the proton polar angle $\theta_{p}$ 
for a simulated $\gamma p \rightarrow \eta^{\prime} p$. Panel b: missing mass spectrum from the recoil proton detection. The black dashed curve shows the effects of selection cuts i) and ii); while the solid blue curve is the result of all preliminary selection cuts i), ii) and iii). Panels c, d and e: Invariant mass spectrum 
from photons (two photons in panels c and e, six photons in panel d) in the BGO calorimeter 
$\it{vs.}$ the missing mass spectrum obtained from the measurement of 
the recoil proton. There are no events in the white area. Panel f: Missing 
mass spectra from the recoil proton measurement after the selection of 
the events in panels c, d and e.\label{Fig1}}\vspace{-0.9cm}
 \end{center}
\end{figure}

The initial event selection, common to all the $\eta^{\prime}$ decay modes, required:\newline
i) at least two photons measured in the ${\it Rugby~Ball}$ for the invariant mass reconstruction;\newline
ii) a tagging energy above $E_{th}$;\newline
iii) a proton detected in the forward TOF wall with polar angle $\theta_{p}$ lying in the acceptance region
shown in Fig. \ref{Fig1}~(a).

The distribution of Fig.~\ref{Fig1} (a) was produced with an upgraded version of the event generator described in \cite{Corvisiero:1994wz}. As we can see, for the photon energies available at GrAAL, the recoil proton is always detected in the forward direction ($\theta_{p}\leq 16^{\circ}$). Moreover, the momentum/energy ratio determined by the two-body kinematics is always below 0.4. We therefore detected non relativistic protons in the forward direction.
In these conditions, the resolution on the  proton momentum for the $\eta^{\prime}$ photoproduction was estimated with a GEANT3\cite{Brun:1987ma} simulation to be  about 2.5\%.

The $\eta^{\prime}$ missing mass calculated from the recoil proton is shown in Fig.~\ref{Fig1}~(b). The effects of the cuts i) and ii) are shown as a black dashed line. The inclusion of cut iii) gave as a result the blue solid line. The $\eta^{\prime}$ peak is clearly visible over a smooth background. This residual background was eventually suppressed by additional constraints on the decay products of the $\eta^{\prime}$ meson.
%
%\begin{figure}[htbp]
%  \begin{center}\vspace{-0.4cm}
%\resizebox{0.5\textwidth}{!}{\includegraphics{fig2a.eps}\includegraphics{fig2b.eps}}
%\put(-233,50){(a)} \put(-105,50){(b)}\\
%\resizebox{0.50\textwidth}{!}{\includegraphics{fig2c.eps}\includegraphics{fig2d.eps}}
%\put(-233,50){(c)} \put(-105,50){(d)}
%\caption{(colour online) Panels a), b) and c): Invariant mass spectrum 
%from photons (two photons in panels a and c, six photons in panel b) in the BGO calorimeter 
%$\it{vs.}$ the missing mass spectrum obtained from the measurement of 
%the recoil proton. There are no events in the white area. Panel (d): Missing 
%mass spectra from the recoil proton measurement after the selection of 
%the events in panels a), b) and c).\label{Fig2}}\vspace{-0.5cm}
%  \end{center}
%\end{figure}

The cleanest decay channel for LAGRAN$\gamma$E is the decay $\eta^{\prime}\rightarrow\gamma\gamma$. The two final-state photons were detected in the ${\it Rugby~Ball}$ and give rise, together with the recoil proton, to the missing mass $\it{vs.}$ invariant mass distribution of Fig.~\ref{Fig1}~(c). This decay mode has a rather small branching ratio ($\simeq 2.20\%$\cite{Agashe:2014kda}) and the number of events collected (3400) did not allow for the extraction of the beam asymmetry with sufficiently good statistics. For this reason, the decay channels involving two pions and one $\eta$ meson were also included in the analysis. The $\eta^{\prime}\rightarrow\pi^{0}\pi^{0}\eta$ decay channel was included by requiring the detection of six photons in the ${\it Rugby~Ball}$ reconstructing the $\eta^{\prime}$ meson invariant mass (Fig. \ref{Fig1}~(d)). For the inclusion of the charged decay channel ($\eta^{\prime}\rightarrow\pi^{+}\pi^{-}\eta$) we required the invariant mass reconstruction from $\eta$ meson decay into two photons (Fig.~\ref{Fig1}~(e)) and two charged tracks in the whole detector, identified as charged pions. All events with extra spurious signals in the detector, charged or neutral, were rejected.

The influence on the missing mass calculated from the recoil proton of the selection on the decay products of the $\eta^{\prime}$ is shown in Fig. \ref{Fig1}~(f). The three missing mass distributions exhibit the same behaviour and the values of the resulting $\eta^{\prime}$ masses are in keeping with the literature\cite{Agashe:2014kda}. At the end of the data reduction, 12121 $\eta^{\prime}$ events are available for asymmetry determination with a residual background, estimated through simulation and mainly due to non-resonant multi-meson photoproduction, of less than 4\%. As the recoil proton angles are the best measured ones, the production angle of the meson in the center-of-mass frame $\theta^{\eta^{\prime}}_{c.m.}$ was calculated from the relevant proton angle $\theta^{p}_{c.m.}$. The angular resolution for $\theta^{p}_{c.m.}$ obtained with this procedure was $\simeq 2^{\circ}$ and no kinematical fit was used to improve it.

\begin{figure}[hbt]
 \begin{center}\vspace{-0cm}
\resizebox{0.45\textwidth}{!}{\includegraphics{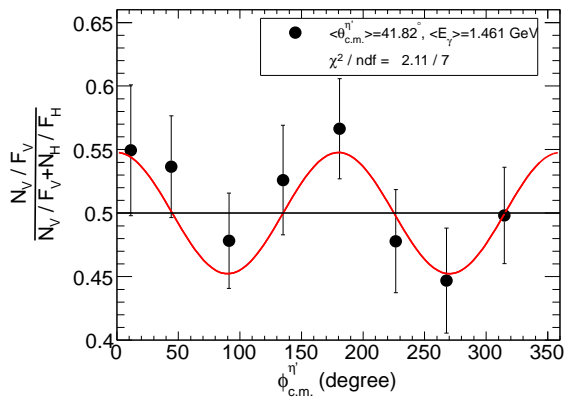}}\vspace{-0.1cm}
\caption{(colour online) Azimuthal distribution at $E_\gamma$ = 1.461 GeV and $\theta^{\eta\prime}_{c.m.}$
= 41.82$^{\circ}$. \label{Fig2}}\vspace{-0.6cm}
 \end{center}
\end{figure}

The selected $\eta^{\prime}$ events were grouped into two energy bins (the first bin is [1.447, 1.475] GeV with centroid 1.461 GeV; the second, with centroid 1.480 GeV, is [1.475, 1.490] GeV), seven angular bins for $\theta^{\eta^{\prime}}_{c.m.}$, and eight for the azimut angle $\phi$.
The beam asymmetry
$\Sigma(E_{\gamma}, \theta^{\eta ^{\prime}}_{c.m.})$ can be calculated by fitting the distribution defined by the following ratio:

\[\frac{N_V/F_V}{N_V/F_V + N_H/F_H} = \frac{1}{2}[1+P(E_\gamma)\cdot \Sigma \cdot cos(2\phi)]\]
\newline
where $N_V$ ($N_H$) and $F_V$ ($F_H$) are the number of events and the total $\gamma$ flux for vertical (horizontal) polarisation states and $P(E_\gamma)$ is the calculated degree of polarisation. Since the kinematics are the same for $H$ and $V$ photons, as is the photon energy distribution, this procedure significantly decreases the systematic errors of the extracted asymmetries, by minimizing the effect of the detection and reconstruction efficiencies. In Fig. \ref{Fig2} we give an example of this azimuthal distribution with the performed fit.

Two sources of systematic errors were considered: i) the possible deterioration of the laser light polarisation on the laser focusing system, with slightly different beam profiles on the target for each polarisation state, and ii) the residual hadronic background. The first error is characteristic of the GrAAL experiment and was established at $\Delta\Sigma$ = 0.02\cite{Bartalini:2007fg}. The second was estimated through determination of two large bins in $\theta^{\eta^{\prime}}_{c.m.}$ ([10,80]$^{\circ}$ and [100,170]$^{\circ}$), and extraction of  asymmetry values from events in the peak of the distribution in Fig. \ref{Fig1}(f) $vs.$ the events belonging to the tails of the same distribution. Peak and tail regions were chosen so that they contain approximately the same number of events, and the results were fairly consistent, with a small decrease in the absolute value of $\Delta\Sigma \sim 0.01$ for the events in the tail regions. Moreover, a MonteCarlo closure test was performed, with trial asymmetry closely reproduced\cite{Mandaglio}. We therefore assumed a total systematic uncertainty $\Delta\Sigma$ = 0.03.

The stability of the results was verified in three alternate ways: i) extraction of the asymmetry with the same large bins in $\theta^{\eta^{\prime}}_{c.m.}$ separately for different stretches of the experiment;
ii) modification of the angular binning, and iii) separate analysis of the subsets of events resulting from neutral or charged decay modes. In all cases, results were satisfactorily stable\cite{Mandaglio}.

\begin{figure}[h]
 \begin{center}\vspace{-0.2cm}
\resizebox{0.45\textwidth}{!}{\includegraphics{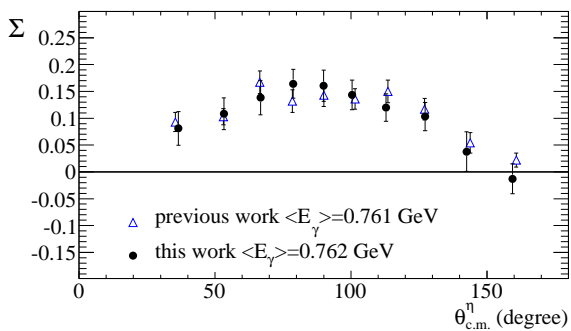}}
\caption{(colour online) $\Sigma$ beam asymmetry for $\eta$ photoproduction obtained in this work (full dots) at $E_{\gamma}$ = 0.762 GeV compared with the results of \cite{Bartalini:2007fg} at $E_{\gamma}$ = 0.761 GeV (open triangles).
 \label{Fig3}}\vspace{-0.8cm}
 \end{center}
\end{figure}

Finally, with the same data set and the same analysis procedure, we extracted the events of the $\eta$ photoproduction process just above the threshold and compared the results with those of \cite{Bartalini:2007fg}. This is a particularly significant test, as the final state detected is exactly the same as with $\eta^\prime$ e.g. $2\gamma$, $6\gamma$ and $2\gamma\pi^{+}\pi^{-}$. Moreover, the data sample in this letter is different from \cite{Bartalini:2007fg}, as was the analysis procedure: in our previous work, only neutral decay channels were considered. The results are shown in Fig. \ref{Fig3} where we compare the values of the asymmetry extracted in this work at 0.762 GeV with the previous GrAAL results at 0.761 GeV. As one can see, the agreement is excellent.

\begin{figure}[hbt]
 \begin{center}
 \vspace{-0.4cm}
\resizebox{0.4\textwidth}{!}{\hspace{-1cm}\includegraphics{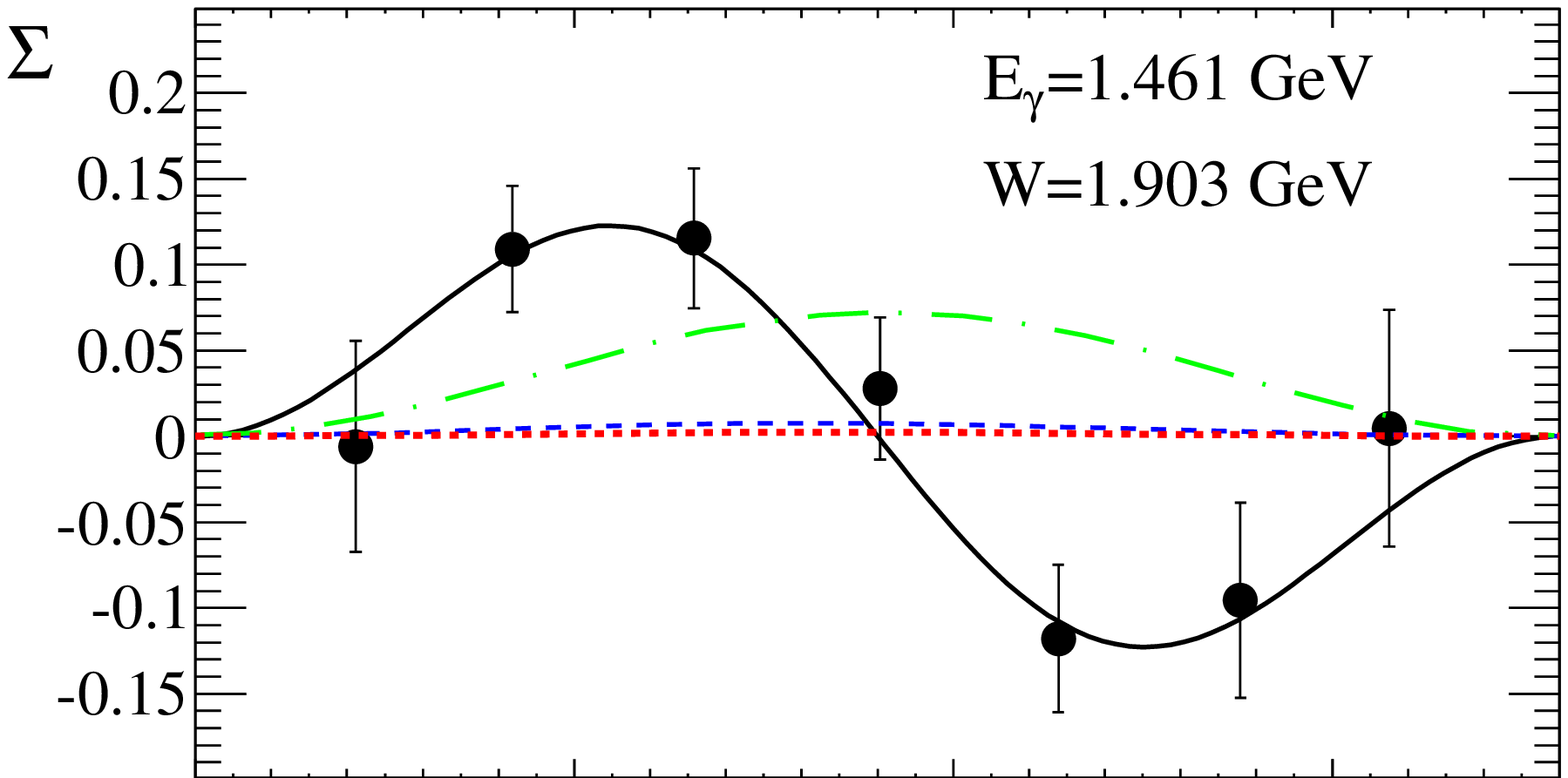}}\vspace{-0.cm}
\resizebox{0.4\textwidth}{!}{\hspace{-1cm}\includegraphics{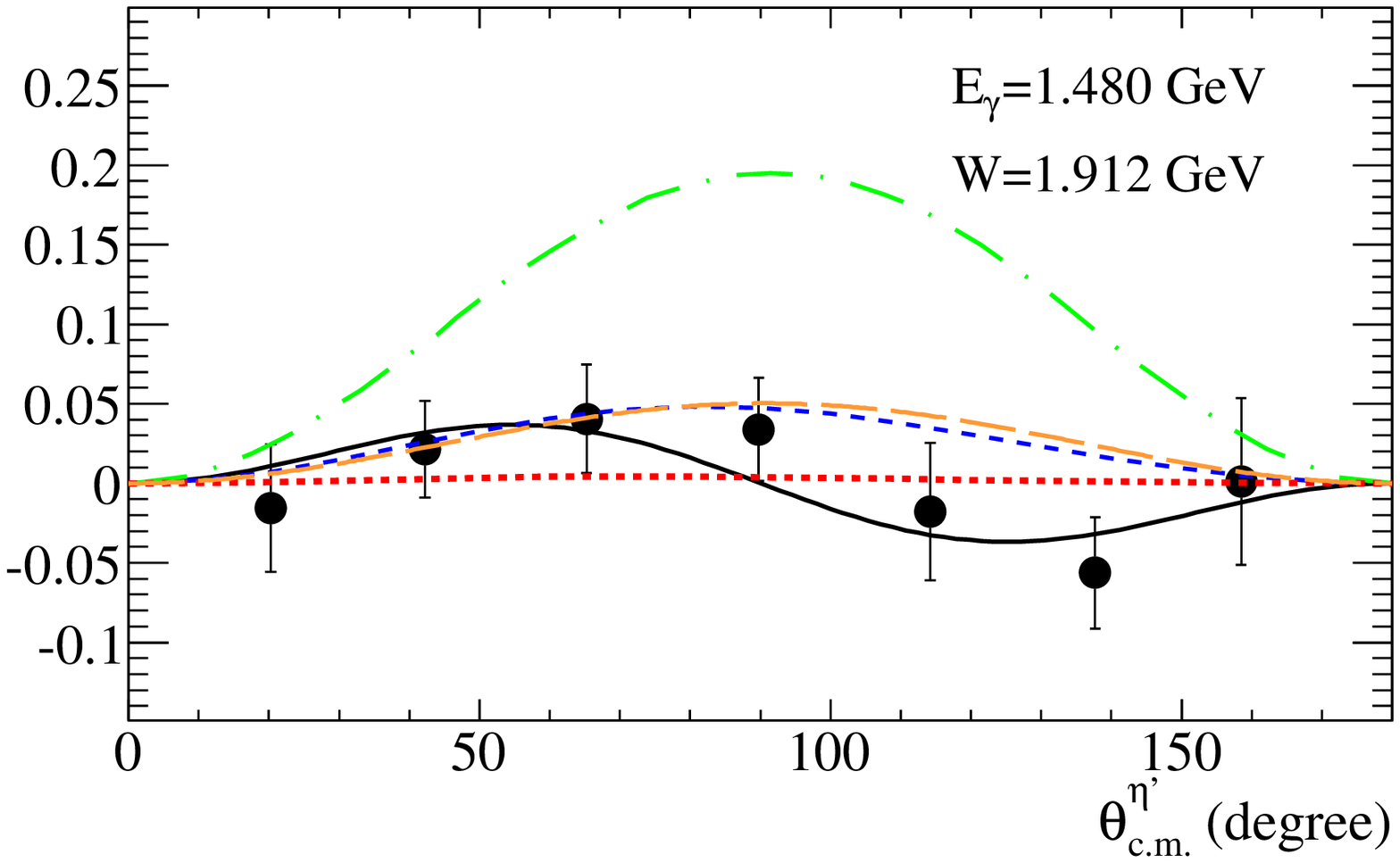}}
\caption{(colour online) $\Sigma$ beam asymmetry at the incoming photon energies of 1.461 and 1.480 GeV
(corresponding to a total center-of-mass energy W of 1.903 and 1.912 GeV respectively)
 as a function of the meson production angle in the center-of-mass system compared to theoretical calculations: red dotted line\cite{Chiang:2002vq}, blue dashed line\cite{Huang:2012xj} green dot-dashed\cite{Tryasuchev:2013fua}, orange long-dashed\cite{Zhong:2011ht}. The solid black line is the result of a fit performed with a function $f(\theta) = a \cdot sin^2(\theta) cos(\theta)$. The fit results for the free parameter are: $ a=0.321\pm 0.063$ at 1.461 GeV and $ a=0.096\pm0.051$ at 1.480 GeV.  \label{Fig4}}\vspace{-0.6cm}
 \end{center}
\end{figure}

The final results of the beam asymmetry $\Sigma$ for the $\eta^{\prime}$ photoproduction process are summarized in Fig. \ref{Fig4} together with the calculations of \cite{Huang:2012xj,Chiang:2002vq,Zhong:2011ht,Tryasuchev:2013fua}. As one can see, the asymmetry is positive at forward angles and negative at backward angles. Moreover, the data indicate a quite strong energy dependence, the effect being more evident at 1.461 GeV, closer to threshold. This behaviour is compatible with a $\sim sin^2(\theta^{\eta\prime}_{c.m.})cos(\theta^{\eta\prime}_{c.m.})$ function, typical of a P-wave D-wave (S-wave F-wave) interference\cite{Drechsel:1992pn,Sandorfi:2010uv}. The existing calculations, whilst providing a reasonable description of the measured cross section, cannot however reproduce these data, especially in the first energy bin ($E_{\gamma}$ = 1.461 GeV corresponding to a total center-of-mass energy of 1.903 GeV) where a change of sign in the asymmetry values around 90$^{\circ}$ for the meson center-of-mass production angle is clearly visible. A slightly better, but still not satisfactory, agreement between data and calculation is obtained at forward angles and at the highest energy bin ($E_{\gamma}$ = 1.480 GeV corresponding to a total center-of-mass energy of 1.912 GeV) in  \cite{Huang:2012xj,Zhong:2011ht}.  We must notice that the theoretical curves presented here are the result of interpolations of the existing models at low energies, and that none of these models contains D-wave or F-wave contributions. It is also important to underline that, in contrast with the conclusions of 
\cite{Crede:2009zzb} for higher energies, at threshold the dynamics of $\eta$ and of $\eta^\prime$ photoproduction processes are clearly different.

These results prove once again that the polarisation degrees of freedom play an essential role in accessing the details of the interaction, and can lead to a better determination of the partial wave contributions and to a better comprehension of the reaction mechanism.

In conclusion, the $\Sigma$ beam asymmetry in the $\eta^{\prime}$ photoproduction was measured at the incoming photon energies of 1.461 and 1.480 GeV by using the highly linearly polarised GrAAL photon beam and the large solid angle LAGRAN$\gamma$E detector. This is the first measurement of this observable for this reaction. The values obtained indicate a P-wave D-wave (S-wave F-wave) interference, the closer to threshold the stronger. Available calculations fail to reproduce the observed behaviour, regardless of the intermediate resonance states involved in the models. 

From the experimental point of view, new measurements with a finer energy binning as well as an extended energy range, would be highly desirable.
%\section*{Acknowledgments}
\\ \\
The authors would like to thank L. Tiator, F. Huang, H. Haberzettl, K. Nakayama, X-H Zhong, Q. Zhao and V. Tryasuchev for kindly providing the results of their models at the energies of this paper, and for helpful discussions. It is a pleasure to thank the ESRF for the reliable and stable operation of the storage ring and the technical staff of the contributing institutions for essential help in the realisation and maintenance of the apparatus.

%\end{acknowledgments}
%\bibliography{etap}

%
% BibTeX users please use
 %\bibliographystyle{}
%\bibliography{etap_2.bib}
%\bibliography{etap_2}

\end{document}